\documentclass[12pt,preprint]{aastex61}

\shorttitle{A SLOWLY POSITIVELY DRIFTING RADIO BURST ...}
\shortauthors{Karlick\'y et al.}
\begin{document}

\title{SOLAR RADIO BURST ASSOCIATED WITH THE FALLING BRIGHT EUV BLOB}

\author[0000-0002-3963-8701]{Marian Karlick\'y}
\affil{Astronomical Institute of the Academy of Sciences of the
Czech Republic, CZ-25165 Ond\v{r}ejov, Czech Republic}
 \email{karlicky@asu.cas.cz}

\author[0000-0002-7565-5437]{Alena Zemanov\'a}
\affil{Astronomical Institute of the Academy of Sciences of the
Czech Republic, CZ-25165 Ond\v{r}ejov, Czech Republic}

\author[0000-0003-1308-7427]{Jaroslav Dud\'{\i}k}
\affil{Astronomical Institute of the Academy of Sciences of the
Czech Republic, CZ-25165 Ond\v{r}ejov, Czech Republic}

\author[0000-0003-0310-1598]{Krzysztof Radziszewski}
\affil{Astronomical Institute, University of Wroc{\l}aw,
51-622 Wroc{\l}aw, ul. Kopernika 11, Poland}

\begin{abstract}
At the beginning of the 4 November 2015 flare, in the 1300 -- 2000 MHz
frequency range, we observed a very rare slowly positively drifting burst. We
searched for associated phenomena in simultaneous EUV observations made by
IRIS, SDO/AIA, Hinode/XRT and in H$_\alpha$ observations. We found that this
radio burst was accompanied with the bright blob, visible at transition region,
coronal, and flare temperatures, falling down to the chromosphere along the
dark loop with the velocity of about 280 km s$^{-1}$. The dark loop was visible
in H$_\alpha$ but disappeared afterwards. Furthermore, we found that the
falling blob interacted with the chromosphere as expressed by a sudden change
of the H$_\alpha$ spectra at the location of this interaction. Considering
different possibilities we propose that the observed slowly positively drifting
burst is generated by the thermal conduction front formed in front of the
falling hot EUV blob.
\end{abstract}


\section{INTRODUCTION}

It is commonly known that a plasma in solar flares is rapidly heated, e.g. by
the magnetic field reconnection. In single loop models of solar flares
or in loops of the flare arcade the plasma may be heated at tops of the loops.
Such a plasma would expand along the loop to its footpoints. This plasma is
collisionless, therefore a question arises if this plasma expansion is a free
expansion of plasma particles without any interactions or the expansion
generates some plasma waves making thus an obstacle for the free expansion.

\cite{Brown79} in their pioneering work answered this question and proposed the
so called thermal conduction front formed between expanding hot and cold
plasmas. Their idea was supported by observations made by
\cite{Farnik83,Rust85,Mandrini96}.

From that time several attempts to simulate the thermal front numerically have
been made. The results of \cite{McKean90a,McKean90b}, based on the 1-D
electrostatic particle-in-cell simulations, did not confirm an existence of the
thermal front. They only showed a free expansion of electrons from the hot
plasma region to the colder one. On the other hand, \cite{Arber09}, using a
Vlasov code, showed that a thermal front trapping of hot electrons can be
really generated.

Recently, \cite{2015ApJ...814..153K} has studied a formation of the thermal
front using a 3-D particle-in-cell fully electromagnetic code. He recognized
the thermal front and importance of inductive effects for its formation. The
thermal front was propagating with the sound speed of the hot plasma. The
thermal front was associated not only with the strong ion-acoustic waves, but
also with the plasma and electromagnetic ones. Just the electromagnetic waves
at the thermal front led to an idea that the thermal front can be observed on
radio waves.

In this article, we present a very rare example of the slowly positively
drifting burst which we propose to be the radio emission of the thermal
conduction front. Such an interpretation is supported by simultaneous EUV and
H$_\alpha$ observations.

\section{OBSERVATIONS}

The studied phenomenon was observed at the beginning of the 4 November 2015
flare. The flare was classified as GOES M3.7 and started at 13:31\,UT, peaked
at 13:52\,UT, and ended at 14:13\,UT and occurred in the active region NOAA
12443, located at this time close to the center of the solar disk.

Global temporal profiles of the flare in \textit{GOES} 1-8 and 0.5-4 \AA~soft
X-rays and their derivatives together with the radio flux observed by the
Ond\v{r}ejov radiometer \citep{JirickaK08} on 3 GHz are presented in
Figure~\ref{fig1}. For global overview of this flare, see movies attached to
the paper by \cite{Li2017}, where they studied evolution of flare ribbons.

\subsection{Radio Spectral Observations}

Figure~\ref{fig2} shows the 800 -- 2000 MHz radio spectrum observed by the
Ond\v{r}ejov radiospectrograph \citep{JirickaK08} at the beginning of the solar
flare in the time interval 13:39:50 -- 13:40:49 UT. As seen here, in the 900 --
1300 MHz frequency range and at 13:39:55 UT the flare radio emission started
with the drifting pulsation structure \citep{Jiricka01}, which drifted with the
frequency drift -18 MHz s$^{-1}$. This drifting pulsation structure is then at
about 13:40:20 UT followed by decimetric type III bursts, indicating the
electron beams propagating upwards through the flare atmosphere. Their
frequency drift is very high, greater than 3 GHz s$^{-1}$.

The most interesting and very rare burst in this radio spectrum is the slowly
positively drifting burst (SPDB). It suddenly started at 1300 MHz, at 13:40:24
UT and drifted towards higher frequencies with the frequency drift of about 115
MHz s$^{-1}$. Its intensity was fading at higher frequencies. The drift
indicates that this burst is generated by the plasma emission mechanism. Thus
assuming that the frequency of SPDB corresponds to plasma frequency then the
density in the radio source is $\geq$ 2.08 $\times$ 10$^{10}$ cm$^{-3}$. Note
that SPDB was observed during the maximum peak of \textit{GOES} 1-8 and 0.5-4
\AA~soft X-rays derivatives (Figure~\ref{fig1}, part b).

After this burst, the Ond\v{r}ejov and e-Callisto \citep{2013EGUGA..15.2027M}
radio spectra show continua with fiber bursts and type II radio burst
(signature of the flare shock) in the 45 -- 80 MHz frequency range at 13:42:00
-- 13:54:00 UT.

\subsection{EUV Imaging Observations}

Using observations made by the Atmospheric Imaging Assembly
\citep[AIA,][]{Lemen12} onboard the Solar Dynamics Observatory
\citep[SDO,][]{Pesnell12}, the X-ray Telescope \citep{Golub2007} onboard of
Hinode satellite \citep{Kosugi07} and the Interface Region Imaging Spectrograph
\citep[IRIS,][]{DePontieu14} we searched for any EUV signature associated with
the slowly positively drifting radio burst. The radio burst was very short
(about 9\,s), so it was challenging to observe its counterpart, considering
cadences of UV/EUV imaging observations. In search of the EUV counterpart we
used the following criteria given by SPDB properties: a) the densities derived
from SPDB should be $\geq$ 2.08 $\times$ 10$^{10}$ cm$^{-3}$, such densities
are commonly observed in EUV \citep{Petkaki2012} and b) the EUV counterpart has
to be observed at time of SPDB and with time duration at least as SPDB, and c)
the positive frequency drift of SPDB indicates that the agent generating SPDB
moves to higher densities, which means in the gravitationally stratified
atmosphere a downward motion. Searching in all flare area at time of SPDB, we
found one distinct bright blob that meets the above criteria. This bright blob
is falling down to the chromosphere and is observed by IRIS, AIA, and XRT
instruments.

Figure~\ref{fig3} contains images from AIA 131\,\AA~and 94\,\AA~filters,
combined with one panel observed in XRT Be\_med filter. We observed the bright
plasma blob (white arrows) falling along the dark loop structure within the
bright flare arcade. The flare emission in AIA 94\,\AA~and 131\,\AA~originates
at temperatures log($T$ [K])\,=\,6.85 and 7.15, respectively but the
contribution form coronal plasma at log($T$ [K])\,$\sim$\,5.45 - 6.05 is
present as well \citep{ODwyer2010,DelZanna2013}. The Be\_med filter is
sensitive to plasma at log($T$ [K])\,$\sim$\,6.5 - 6.9 \citep{ODwyer2014} while
the coronal emission below log($T$ [K])\,$\lesssim$\,6.1 is suppressed
\citep{Narukage2014}. Simultaneous observations in AIA 94, 131\,\AA~and XRT
Be\_med filters point to a hot component within the blob, likely with log($T$
[K])\,$\geq$\,6.9.

Figure~\ref{fig4} shows the transition region and coronal counterpart of the
hot emission in Figure~\ref{fig3}. This figure contains the IRIS
1400\,\AA~slit-jaw images (SJI) combined with AIA 211\,\AA~and AIA
304\,\AA~filter. There are no IRIS spectra of the blob since the slit did not
reach it.

The 1400\,\AA~IRIS filter is dominated by transition region lines of
\ion{Si}{4} formed at log($T$ [K])\,$\sim$\,4.9 \citep{Dudik14} but contains
also photospheric continuum emission. The AIA 304\,\AA~filter is dominated by
optically thick \ion{He}{2} Ly-$\alpha$ line at 303.8\,\AA~formed at around
log($T$ [K])\,$\approx$\,4.9 \citep{Dere1997, DelZanna2015}.

The emission at coronal temperatures should be seen by the AIA filters 171,
193, and 211\,\AA. The 171 and 193\,\AA~observations were overexposed, thus we
present AIA 211\,\AA~filter observations only. The emission seen in this
passband is not entirely understood \citep{DelZanna2011, DelZanna2013}. The
coronal contribution of \ion{Fe}{14} 211.3\,\AA~line at log($T$ [K])\,=\,6.3
can be present also under the flare conditions but the most significant
contribution (in flare) comes from continuum emission \citep{ODwyer2010}. Thus
a coronal emission is very likely present within the blob.

At transition region temperatures ribbons are well visible. The blob is also
visible in AIA 304\,\AA~and 211\,\AA~observations (white arrows) and in IRIS
1400\,\AA~SJI. Yellow arrows in the first two panels of 304\,\AA~images show
the position of dark loop structure along which the bright blob falls.

All these observations show that the falling bright blob has a
multi-temperature structure spanning within two orders of magnitude in
temperature.

We tried to estimate velocity of the falling blob in the plane of image from
AIA filters in Figures~\ref{fig3}~and~\ref{fig4}. The center position of the
blob was marked manually at each image. From these positions and the cadence we
estimated velocity, which is 279\,(+31/-33) km s$^{-1}$ during 13:40:06 --
13:40:24\,UT, when the blob moved through the largest arc of the dark loop.

The accompanying movie in 304\,\AA~filter shows that during 13:39:00\,UT -
13:40:00\,UT, the SPDB connected falling blob is preceded by other, but less
bright blobs falling along the wider dark loop structure but at slightly
different trajectories, perhaps along neighboring dark threads. They move on
both sides of the dark loop and are immediately followed by the middle one,
connected with SPDB. After this blob the dark loop disappeared. Only the blob
connected with SPDB was visible in XRT Be\_med and Al\_thick filters indicating
the highest temperature in this blob. This type of blobs resembles the
flare-driven coronal rain clumps \citep{Scullion16}. However, our blobs are
observed at the impulsive phase of the flare, contrary to the rain clumps that
are observed at the end of the flare \citep{Scullion16}. The velocity of our
blob is faster than that of the rain clumps (several tens km s$^{-1}$) and the
blob temperature is higher than in rain clumps ($\sim$ 22000 K - $\sim$ 1 MK).
In this sense, our observed blobs are unique, and it is unlikely that they
occur due to coronal rain.

\subsection{H$_\alpha$ Observations}

The studied flare was also observed in the H$_\alpha$ (6562.8\,\AA) by
the Multi-channel Subtractive Double Pass (MSDP) imaging spectrograph and Large
Coronograph (LC) at the Bia{\l}k{\'o}w Observatory, Poland \citep{Mein91,
Romp94}.

On November 4th, 2015 LC worked with artificial moon removed, enabling
observations of the solar surface. Between 13:12\,UT and 14:15\,UT, 110 scans
of NOAA~12443 were collected. The 2D spectra-images obtained from LC-MSDP
system have pixel size $\sim$0.5'' and the spatial resolution of LC is limited
by seeing conditions to $\approx$~1''. The nine-channel MSDP prism-box enables
recording spectra-images at nine positions across the H$_\alpha$ line profile,
with a total range of $\pm$ 1.6\,\AA. After a standard dark current and
flat-field reduction, two-dimensional, quasi-monochromatic images with pixel
band-width of 0.06\,\AA, separated by 0.2\,{\AA} up to $\pm$ 1.2\,{\AA} from
the H$_\alpha$ line center are restored \citep{Rad06, Rad07}. Thus the
H$_\alpha$ profile in the range of $\pm$1.2\,{\AA} from the line center is
available for each pixel within the FOV.

We have studied H$_\alpha$ observations to look for any optical counterpart of
radio/EUV event described in previous sections. Figure~\ref{fig5} shows two
MSDP images in the center of H$_\alpha$ line at times close to that of SPDB. In
the first panel the white arrow shows the position of the bright plasma blob
visible at the same time (13:40:12\,UT) in AIA 94\,\AA. There is no counterpart
of the blob visible in H$_\alpha$ in this position at this time. Instead, a
dark loop structure in H$_\alpha$ (log($T$ [K])\,$\approx$\,4) is already
visible since about 13:35\,UT (see online movie in H$_\alpha$). It quickly
evolved from a wider structure into narrow fibre at about 13:40:12\,UT
(Figure~\ref{fig5}, the first panel), and finally disappeared. A bright knot
appeared in its place at about 13:40:42\,UT (Figure~\ref{fig5}, the second
panel), i.e., at the time when the UV/EUV blob reached the ribbon.

The H$_\alpha$ profiles (Figure~\ref{fig5}, last panel) were taken from the
same position, i.e. from the central part of the dark loop structure at
13:40:12\,UT and the bright knot at 13:40:42\,UT, marked by the white crosses
in Figure \ref{fig5}. The dark loop structure at 13:40:12\,UT is characterized
by a small redshift corresponding to velocity of +11~km s$^{-1}$. After 30s, in
the same position, the bright knot was visible, whose H$_\alpha$ profile was
significantly increased. In red wing there is clear local maximum of intensity,
which corresponds to plasma moving downward with velocity of +41~km s$^{-1}$ -
i.e. almost three times more than earlier. Additionally, in blue wing, the
smaller local maximum was visible, which implied the second component of upward
moving plasma with velocity of -35~km s$^{-1}$. Such a profile
suggests presence of both direction of plasma motion in the bright knot but
with dominating downward component.

\section{DISCUSSION AND CONCLUSIONS}

The slowly positively drifting burst (SPDB) was observed at the beginning of
solar flare and its drift (115 MHz s$^{-1}$) was much smaller than that of
following decimetric type III bursts (several GHz s$^{-1}$). Because type III
bursts are generated by electron beams with the velocity of about one third of
speed of light, SPDB has to be generated by an agent having much smaller
velocity.

The presented EUV observations show that at the time of SPDB the bright blob
was propagating along the dark loop, visible in H$_\alpha$, downwards to the
chromosphere. An interaction of this falling blob with the chromosphere was
expressed by a sudden change of the H$_\alpha$ line spectrum, which shows a
strong plasma heating and enhanced plasma velocities.

The fall velocity of the EUV blob is estimated as about 280 km s$^{-1}$. It is
much smaller than the Alfv\'en speed at such atmospheric altitudes (about 1000
km s$^{-1}$), but greater than the velocities of flare-driven coronal rain
clumps. As argued in Section 2.2, our blob is not due to the coronal rain
phenomenon.

Based on these observations, firstly, we considered the model with the magnetic
island (plasmoid) moving downwards. Plasmoids are observed in the
flare impulsive phase. But this explanation is not very probable because the
plasmoid has to move along the current sheet \citep{2008A&A...477..649B} and no
indications of the straight or distorted current sheet were found in and around
the loop where the EUV blob was moving. Moreover, such a plasmoid would be
magnetically open in the direction of the magnetic field lines, i.e., in the
direction of the loop and thus it would need some structure like the thermal
front to be thermally isolated from cold plasma. We also considered some
dissipation process along the loop with the twisted magnetic field lines. But
in this case the dissipation spreads with the Alfv\'en velocity that is higher
than the observed one. The problem with the thermal isolation in the direction
of the loop axis would be the same as in the plasmoid case. Thus on the results
obtained by numerical simulations \citep{2015ApJ...814..153K}, we propose that
SPDB is generated by the thermal conduction front formed in front of the
falling bright and hot blob. Namely, the thermal front can generate the
electromagnetic (radio) waves by plasma emission mechanism.

\begin{figure*}
\begin{center}
\includegraphics[width=8cm]{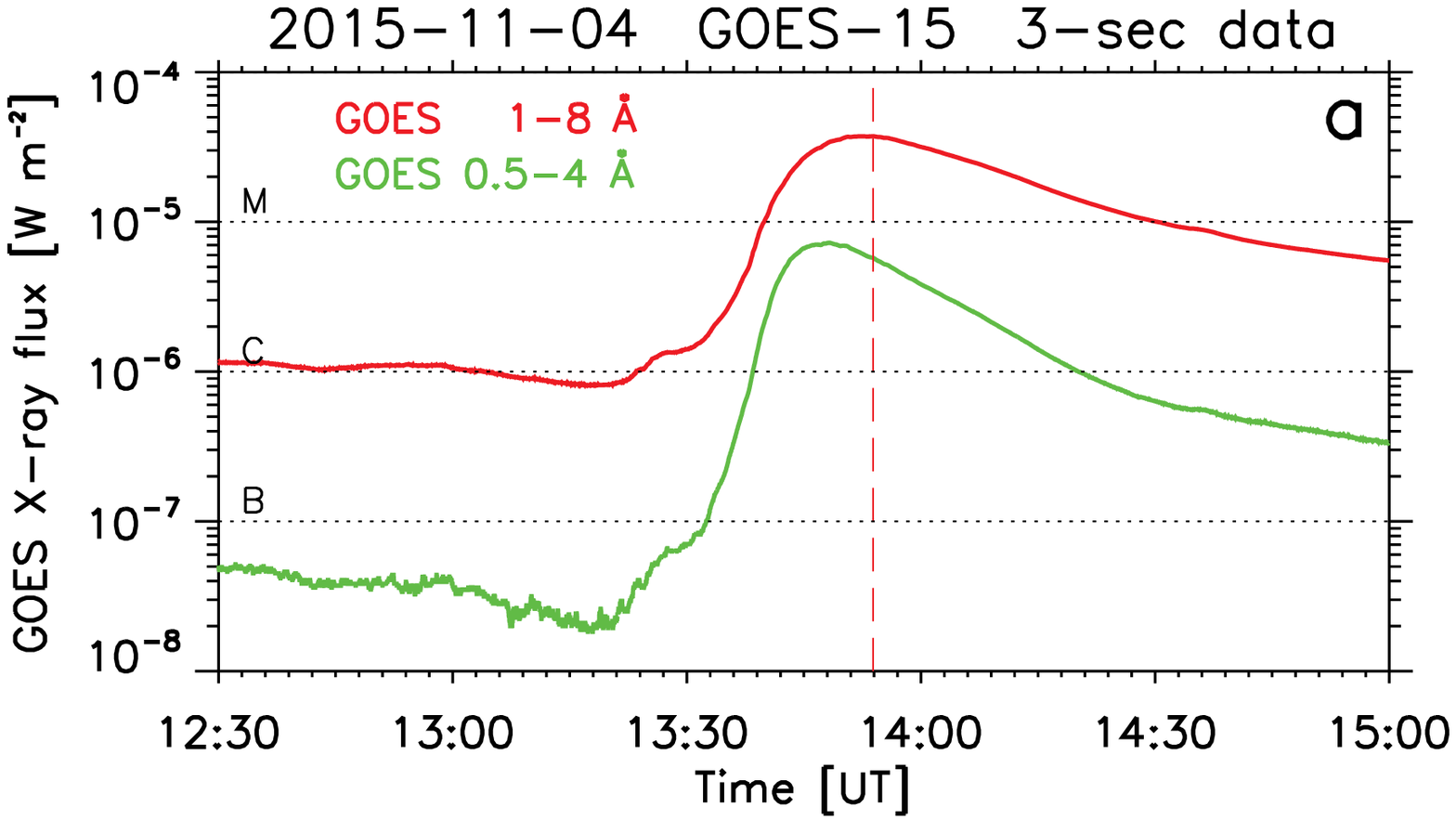}
\includegraphics[width=8cm]{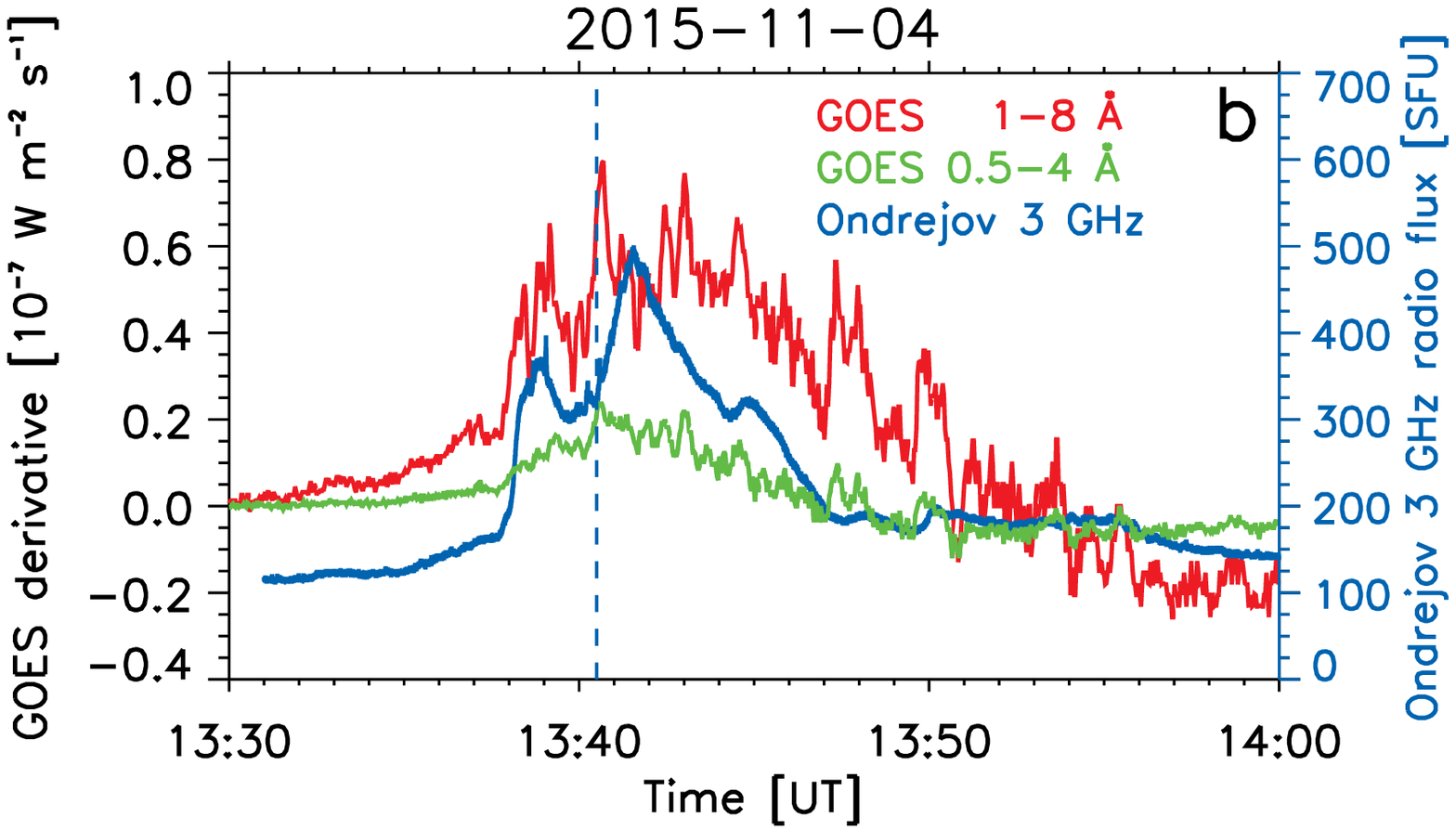}
\end{center}
    \caption{\textit{a) GOES} soft X-ray fluxes at 1-8 \AA (red) and 0.5-4 \AA (green) during the M3.7. The red dashed vertical line designates a maximum of the flare.
    b) Time derivative of GOES flux in both channels (red and green) and radio flux observed by the Ond\v{r}ejov radiometer
    at 3000 MHz (blue). The blue dashed vertical line shows the observational time of the slowly positively drifting burst (SPDB) (see Figure~\ref{fig2}).}
    \label{fig1}
\end{figure*}

\begin{figure}
\begin{center}
\includegraphics*[width=18.0cm]{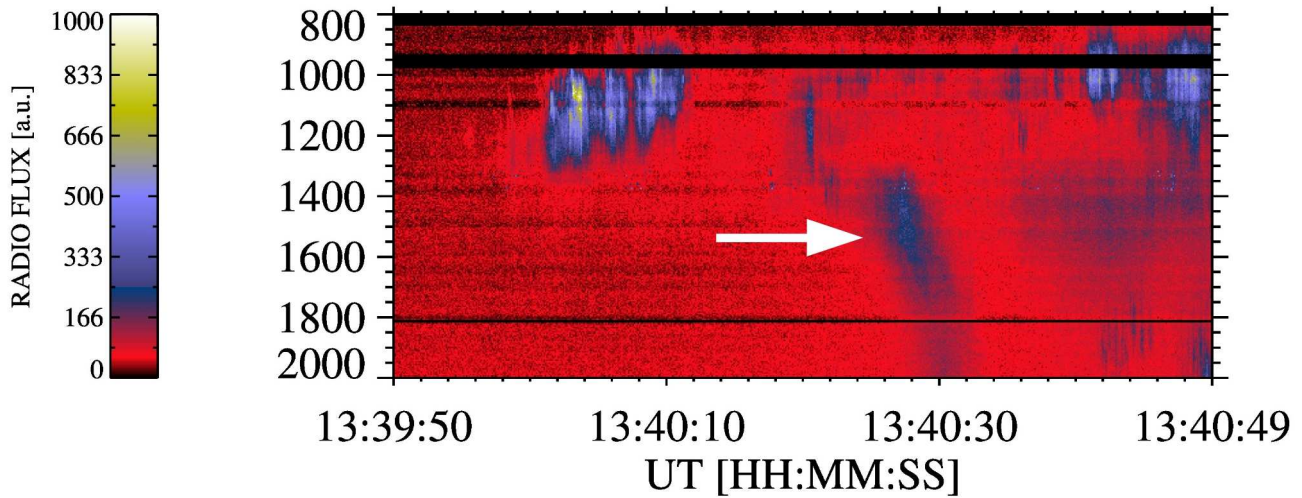}
\end{center}
    \caption{The 800-2000 radio spectrum observed during the 4 November 2015 flare
by the Ond\v{r}ejov radiospectrograph. The arrow shows the SPDB.}
\label{fig2}
\end{figure}

\begin{figure*}
        \centering
        \includegraphics[width=5.27cm,clip,viewport= 0 50 350 305]{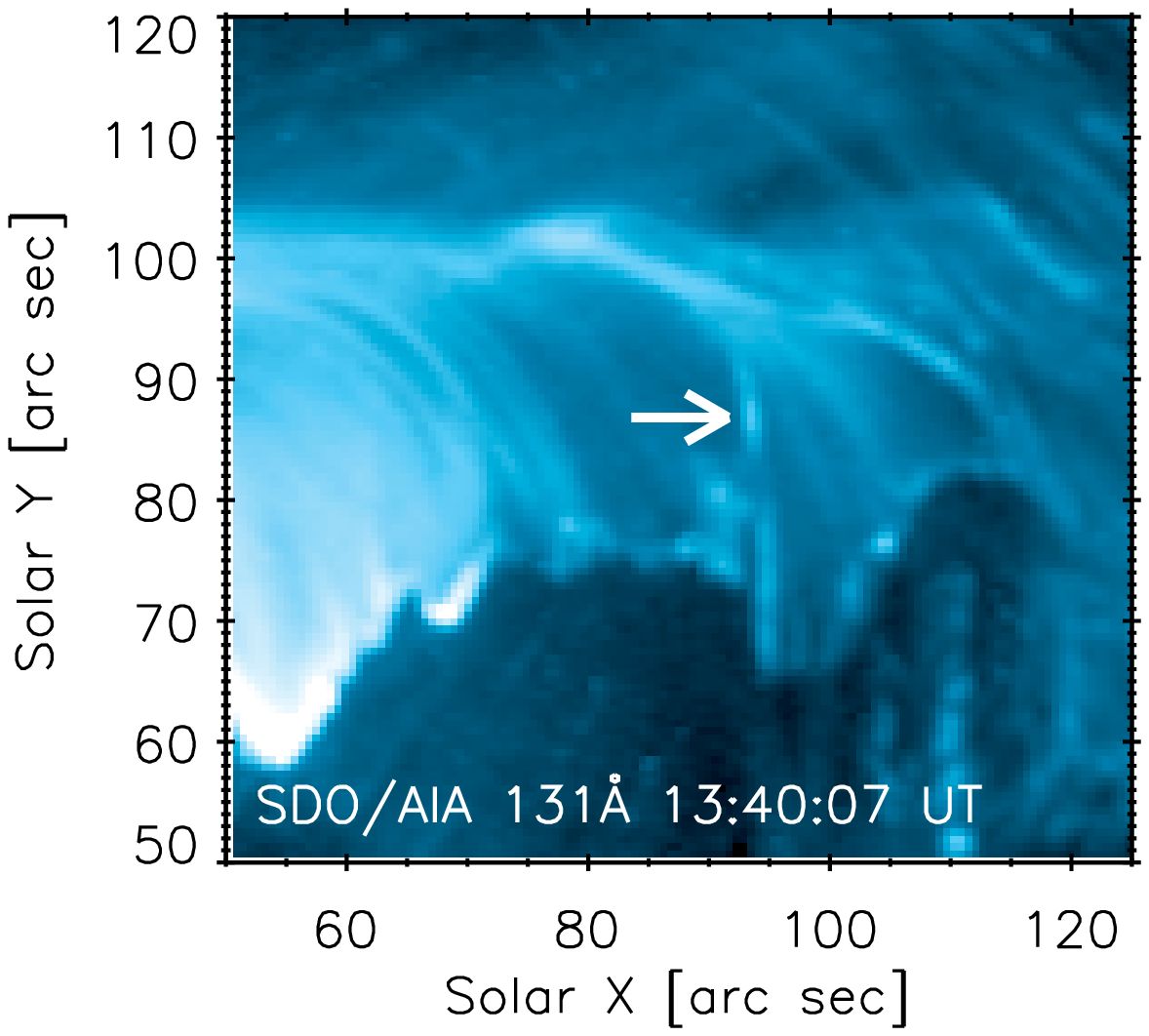}
        \includegraphics[width=4.11cm,clip,viewport=77 50 350 305]{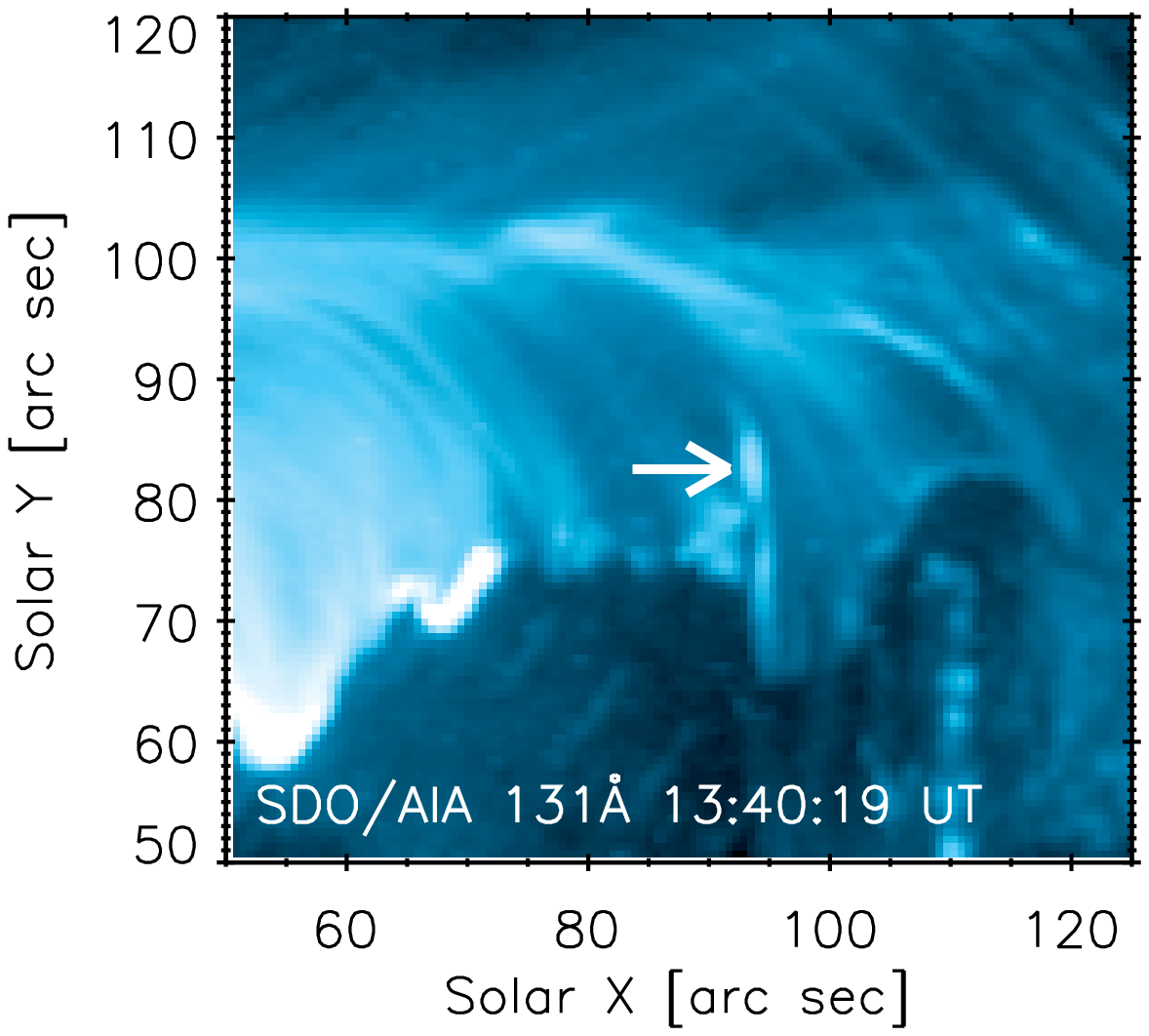}
        \includegraphics[width=4.11cm,clip,viewport=77 50 350 305]{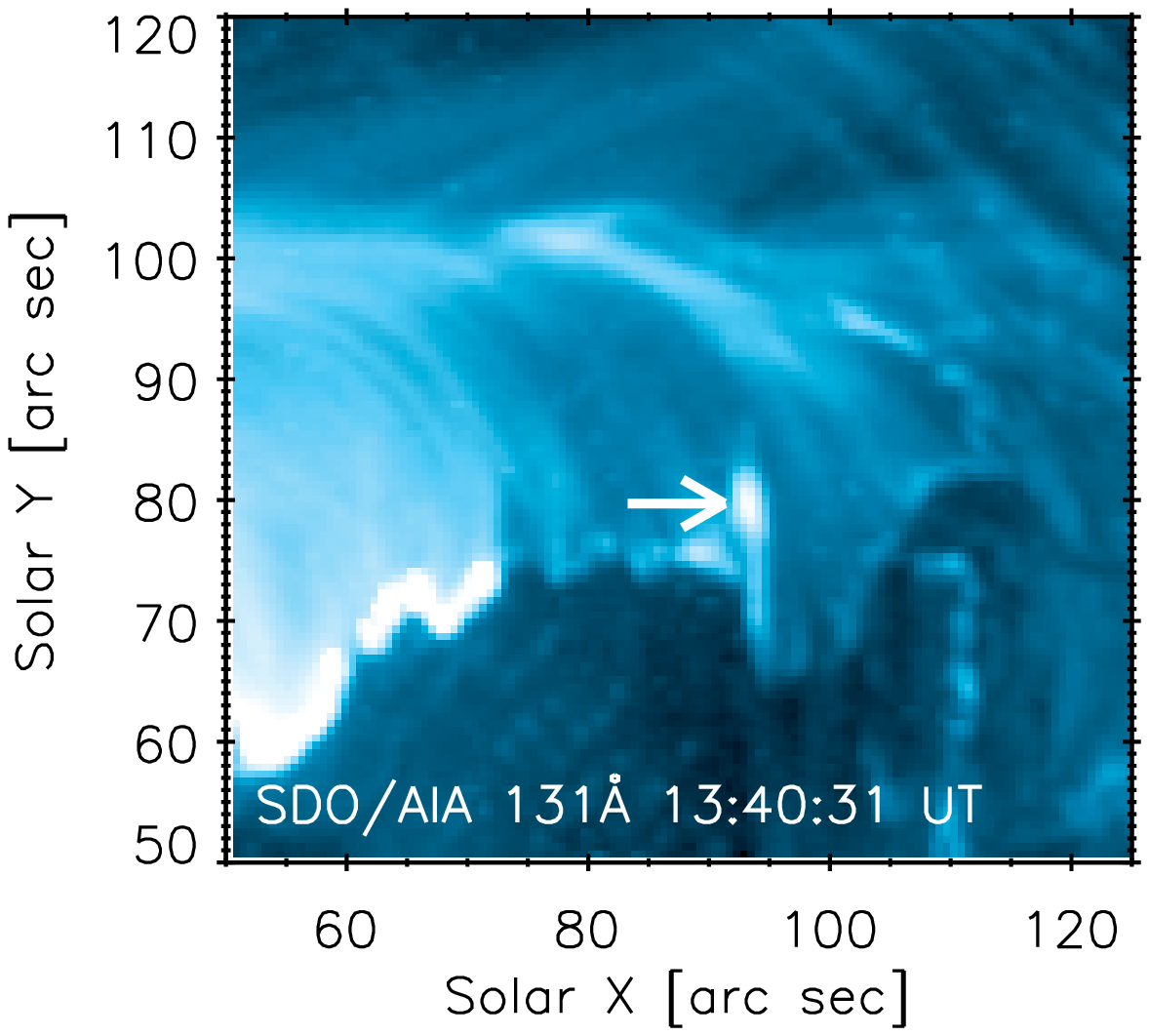}
        \includegraphics[width=4.11cm,clip,viewport=77 50 350 305]{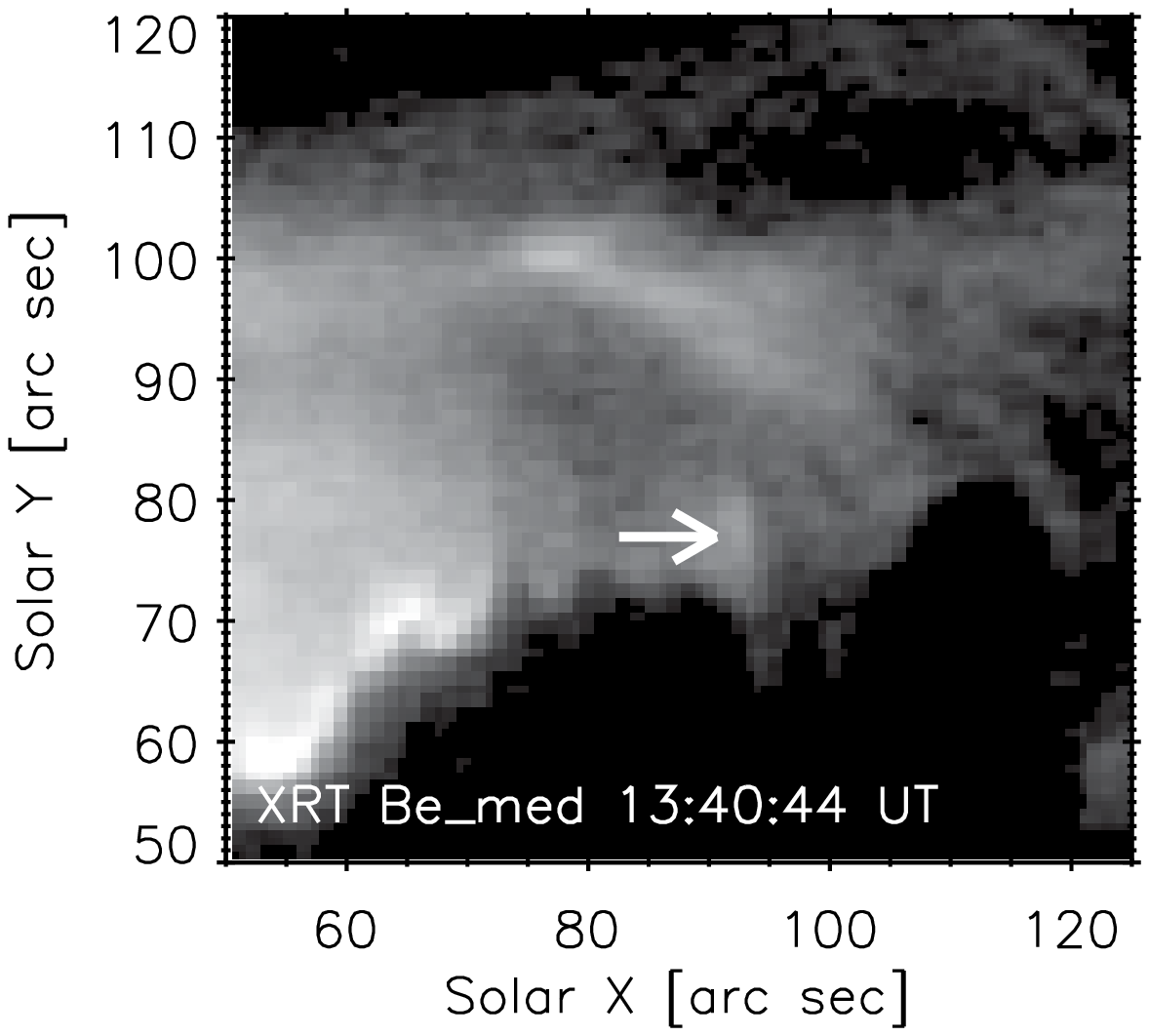}

        \includegraphics[width=5.27cm,clip,viewport= 0  0 350 305]{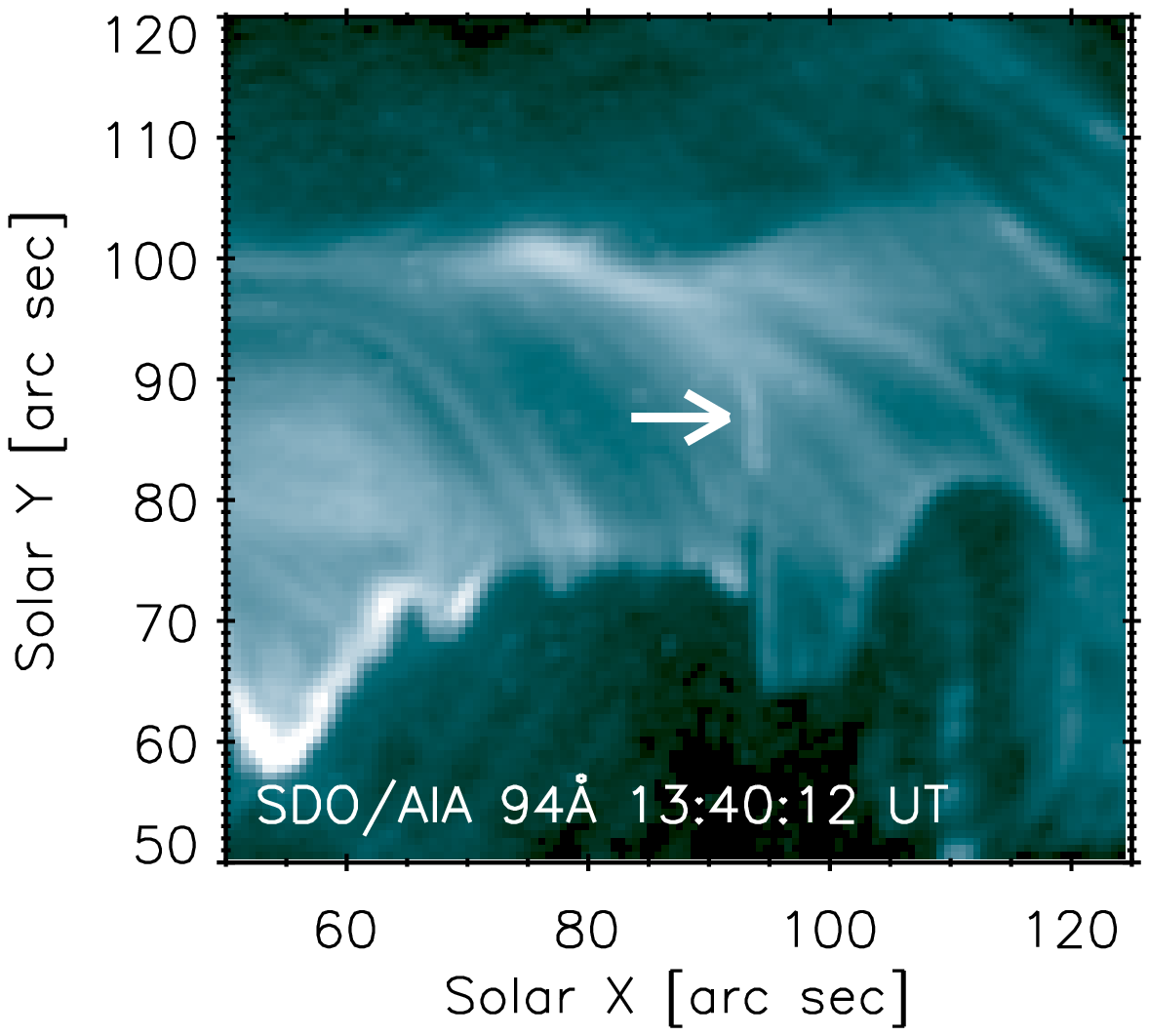}
        \includegraphics[width=4.11cm,clip,viewport=77  0 350 305]{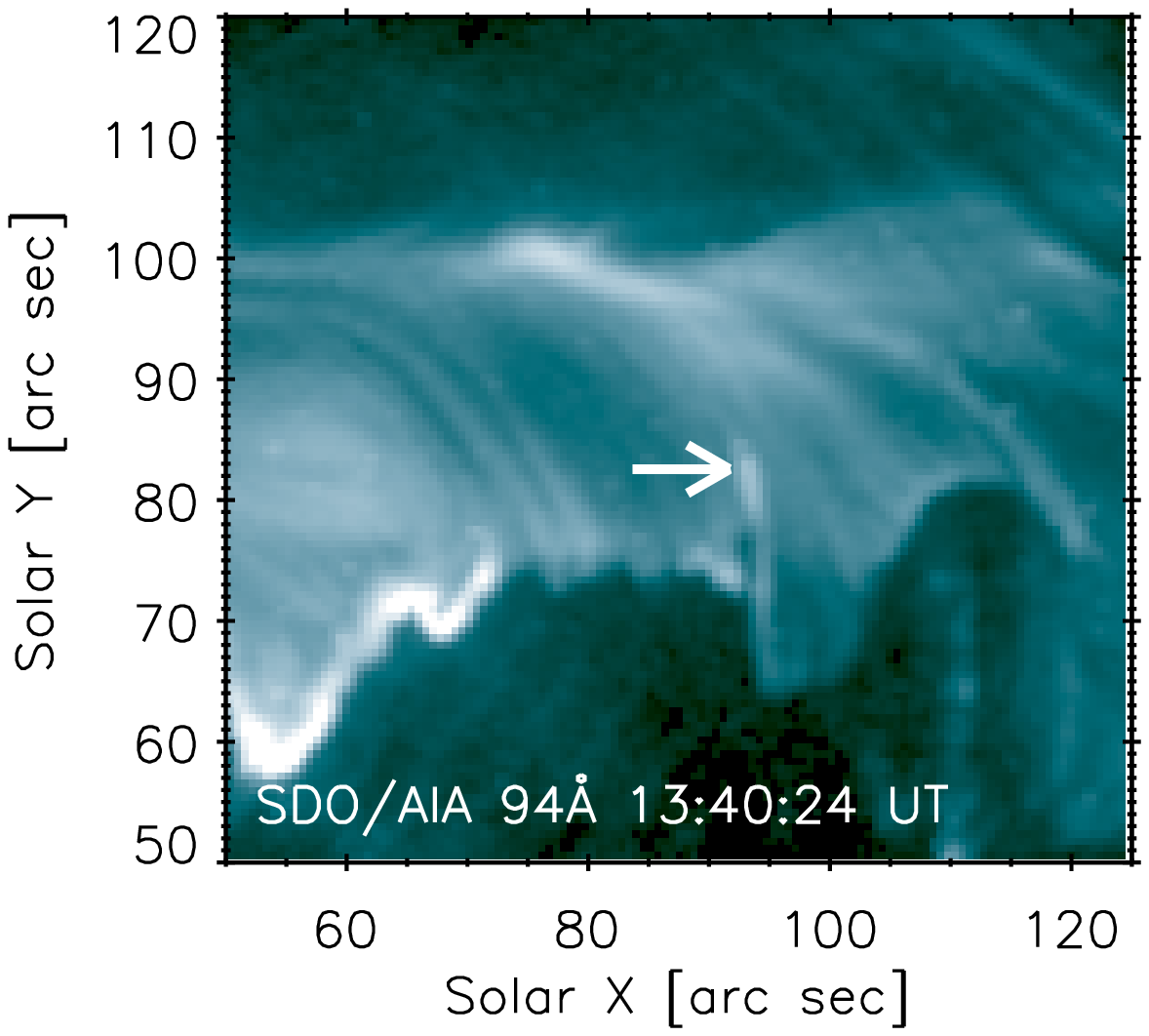}
        \includegraphics[width=4.11cm,clip,viewport=77  0 350 305]{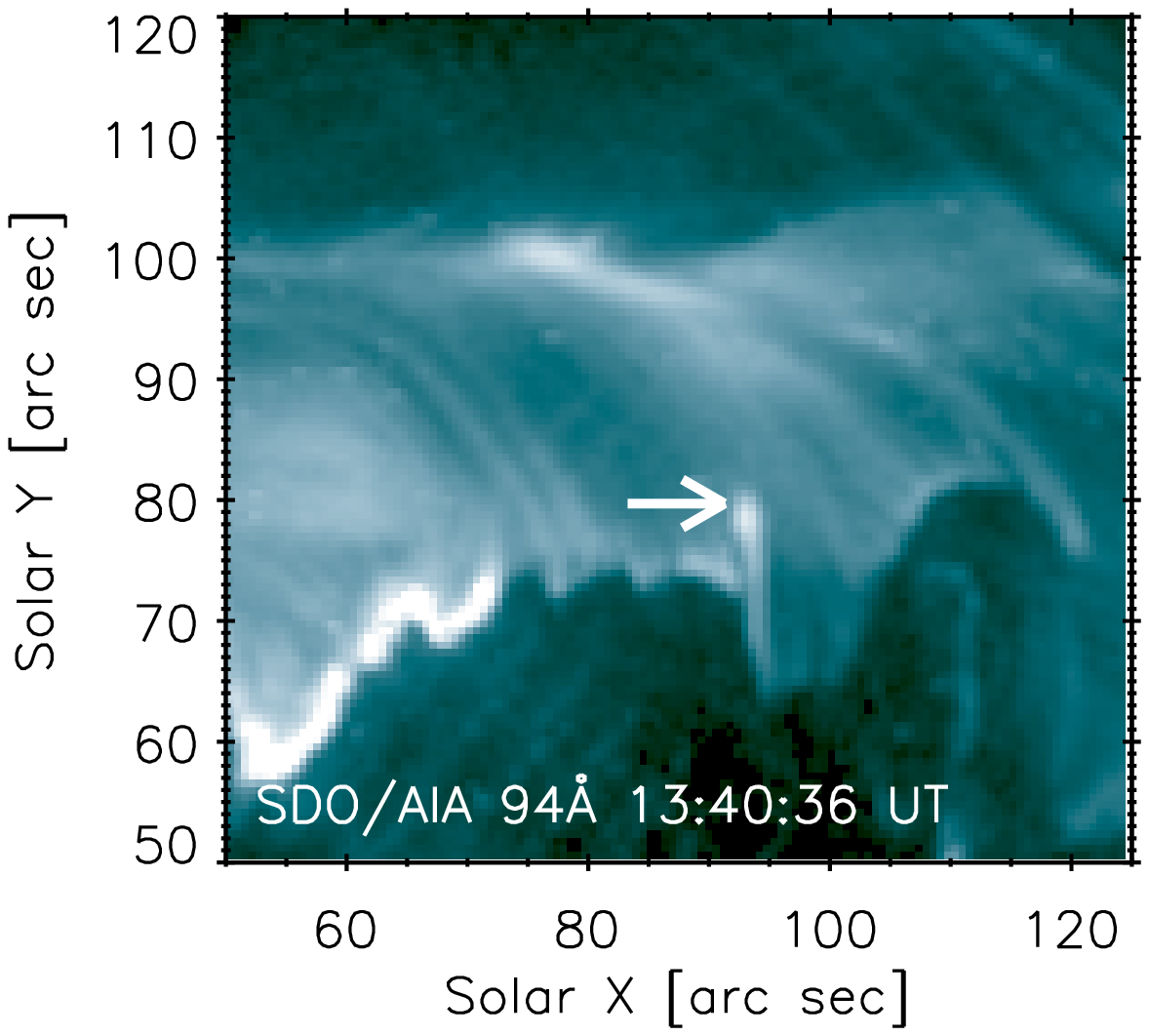}
        \includegraphics[width=4.11cm,clip,viewport=77  0 350 305]{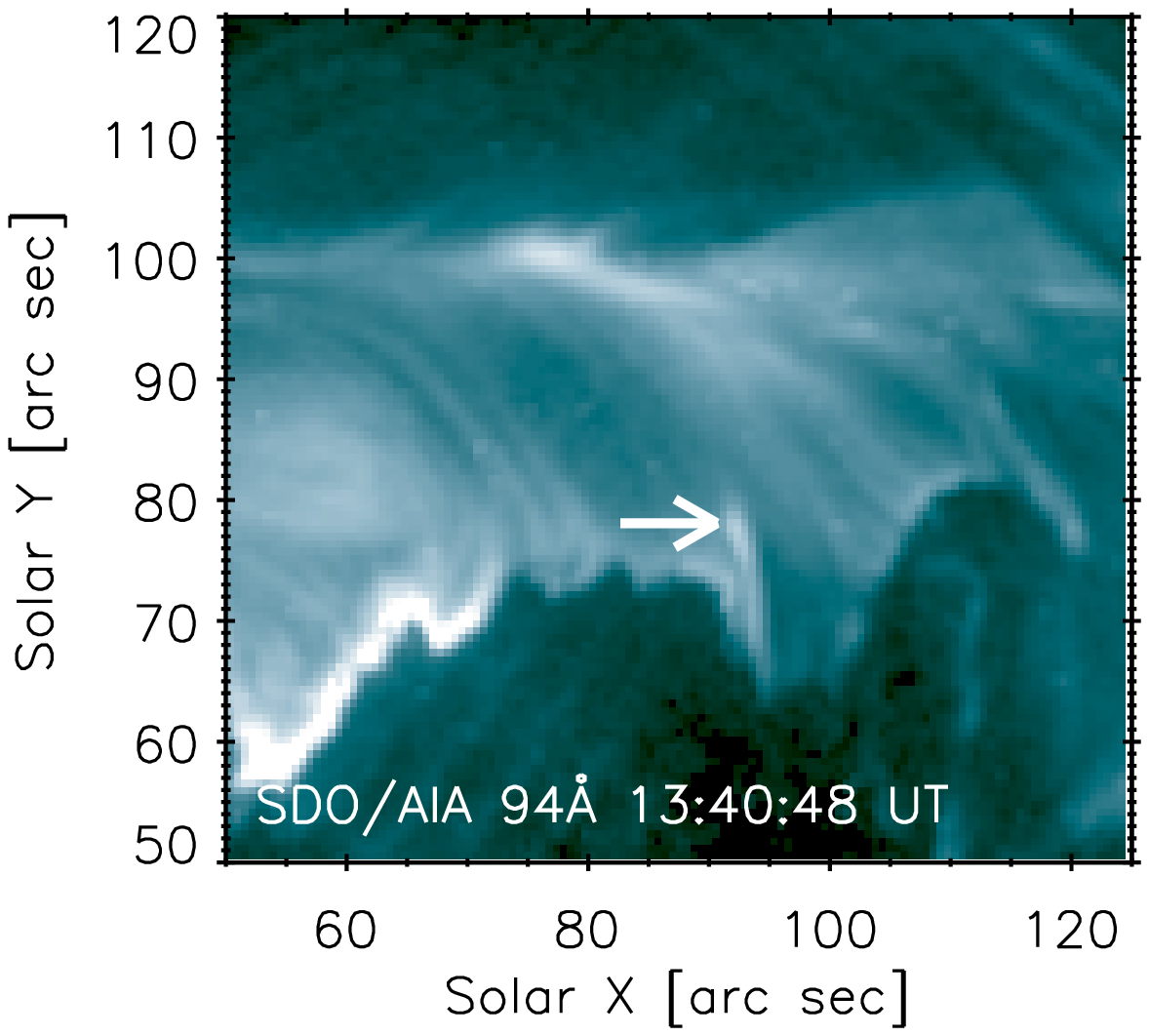}

\caption{Motion of plasma blob along the dark loop structure as seen in SDO/AIA
and Hinode/XRT filters: 131\,\AA~and Be\_med (top row) and 94\,\AA~(bottom
row). White arrows track the motion of the blob.} \label{fig3}
\end{figure*}

\begin{figure*}
        \centering
        \includegraphics[width=5.27cm,clip,viewport= 0 50 350 305]{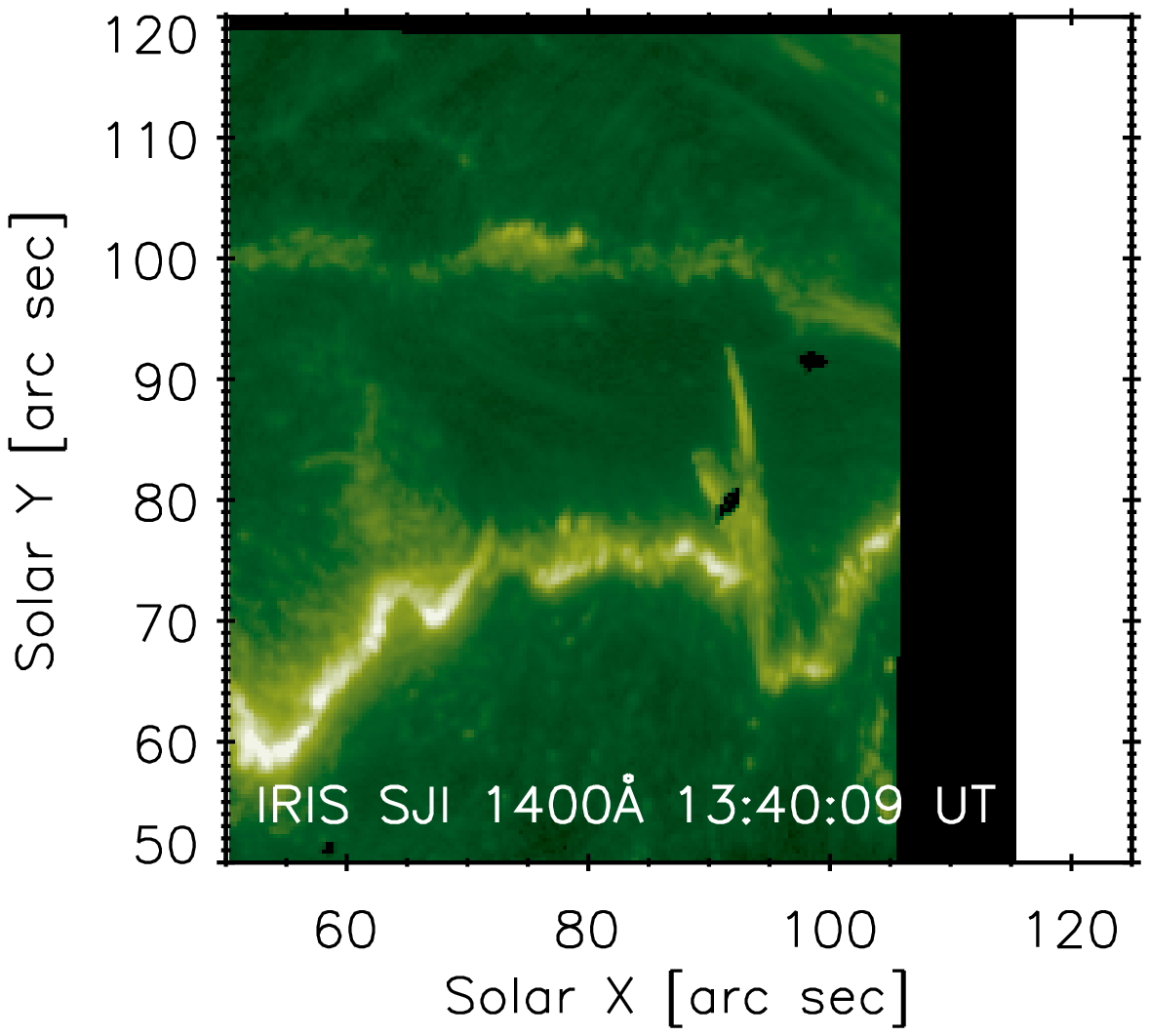}
        \includegraphics[width=4.11cm,clip,viewport=77 50 350 305]{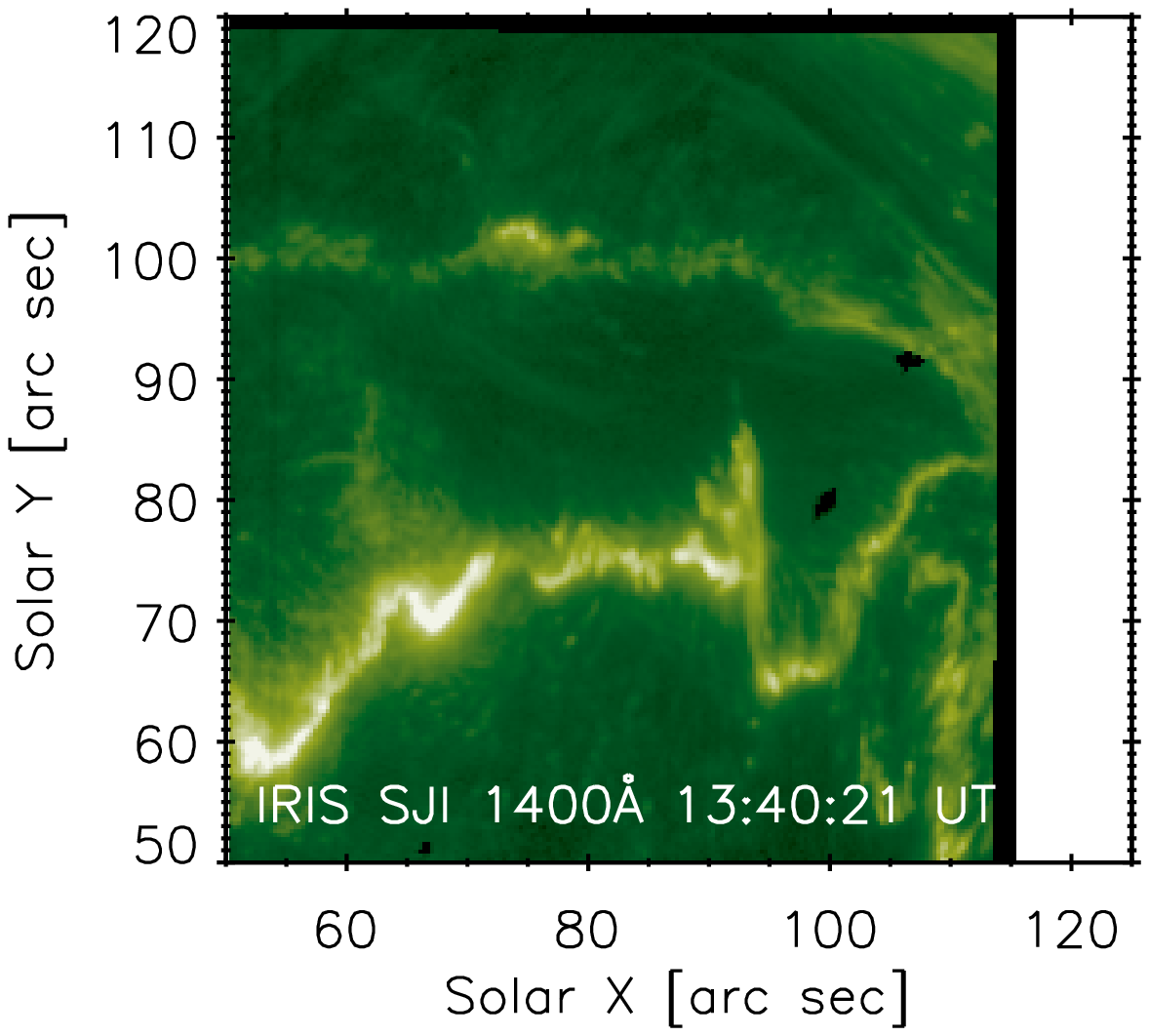}
        \includegraphics[width=4.11cm,clip,viewport=77 50 350 305]{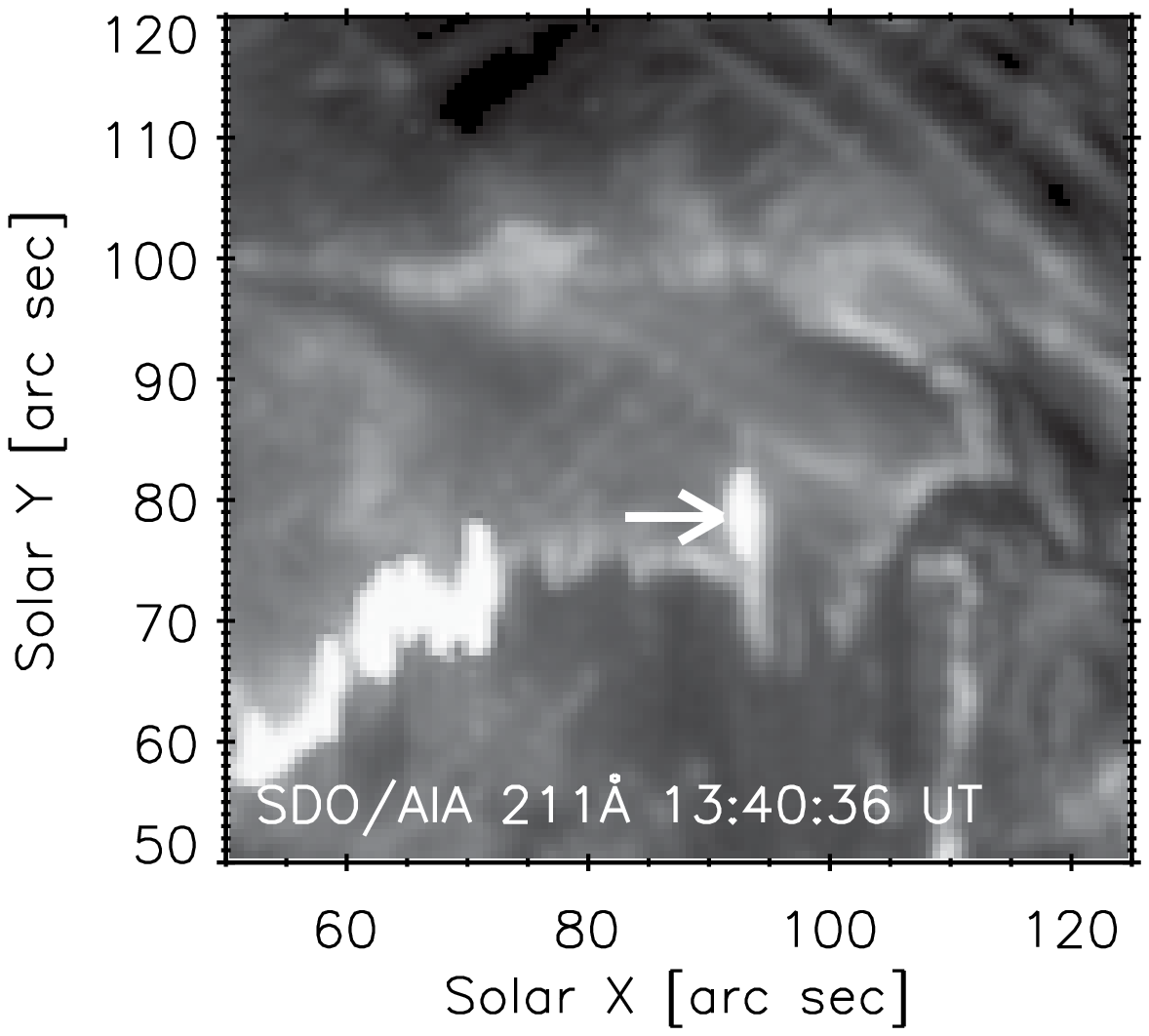}
        \includegraphics[width=4.11cm,clip,viewport=77 50 350 305]{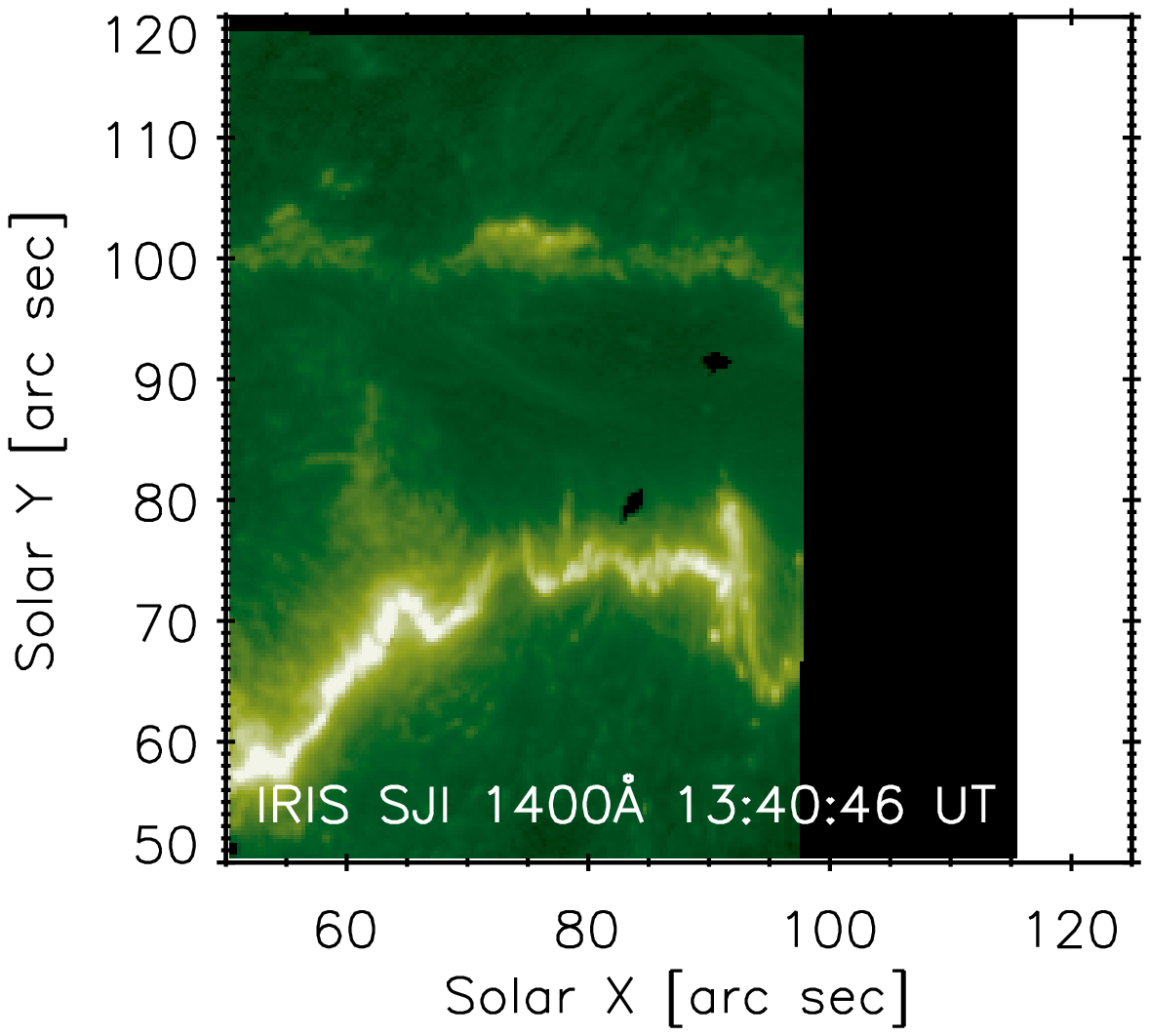}

        \includegraphics[width=5.27cm,clip,viewport= 0  0 350 305]{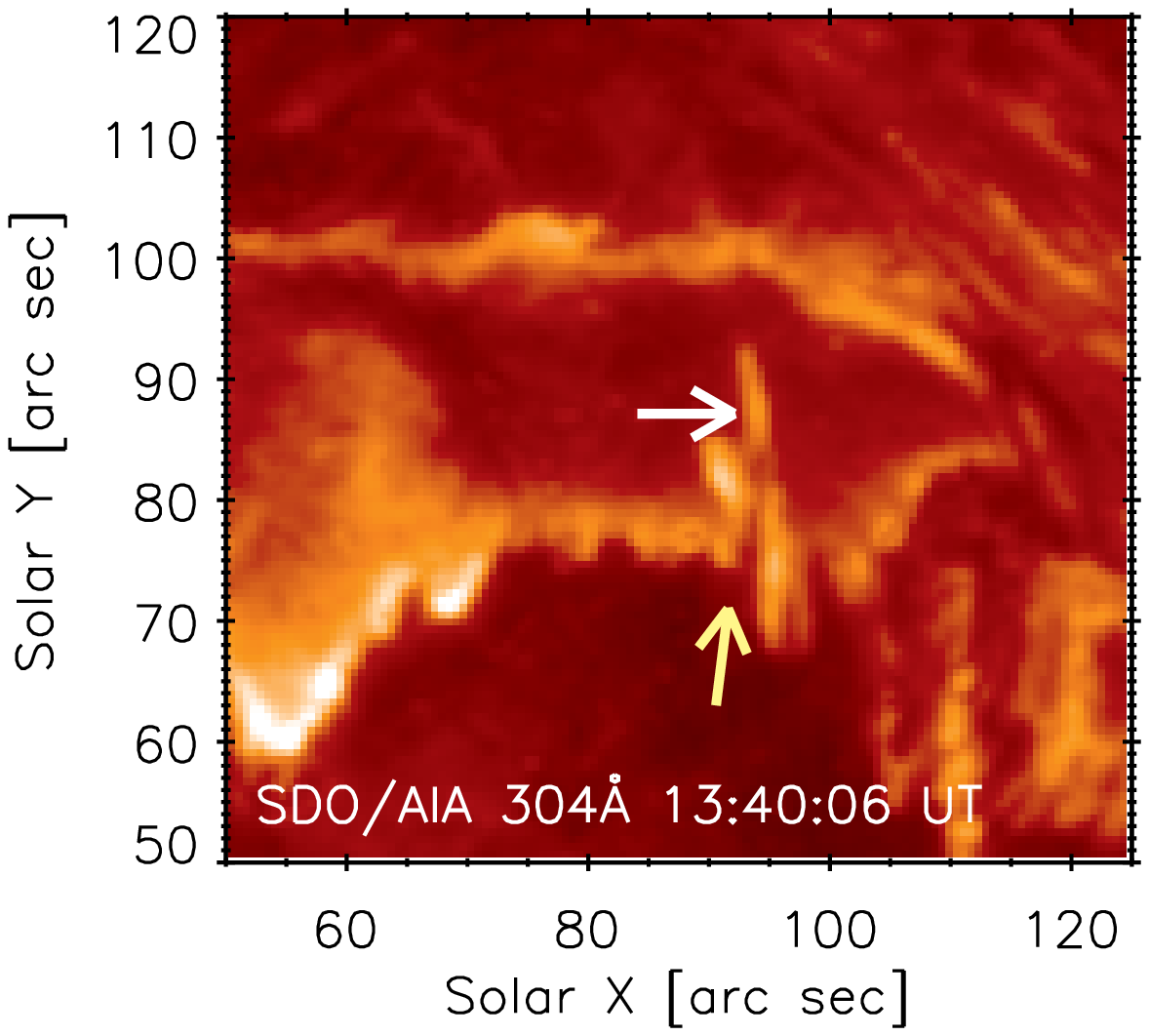}
        \includegraphics[width=4.11cm,clip,viewport=77  0 350 305]{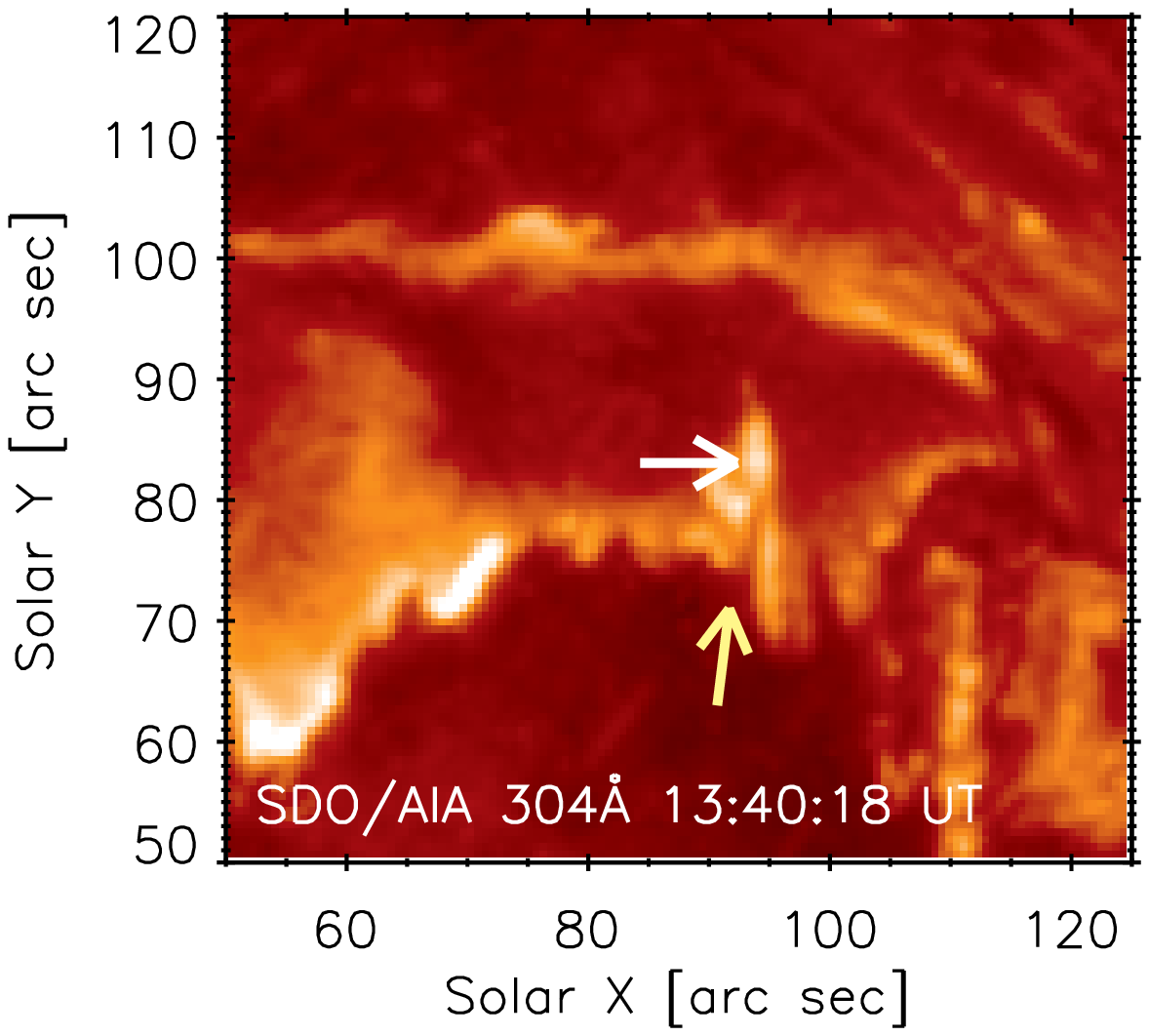}
        \includegraphics[width=4.11cm,clip,viewport=77  0 350 305]{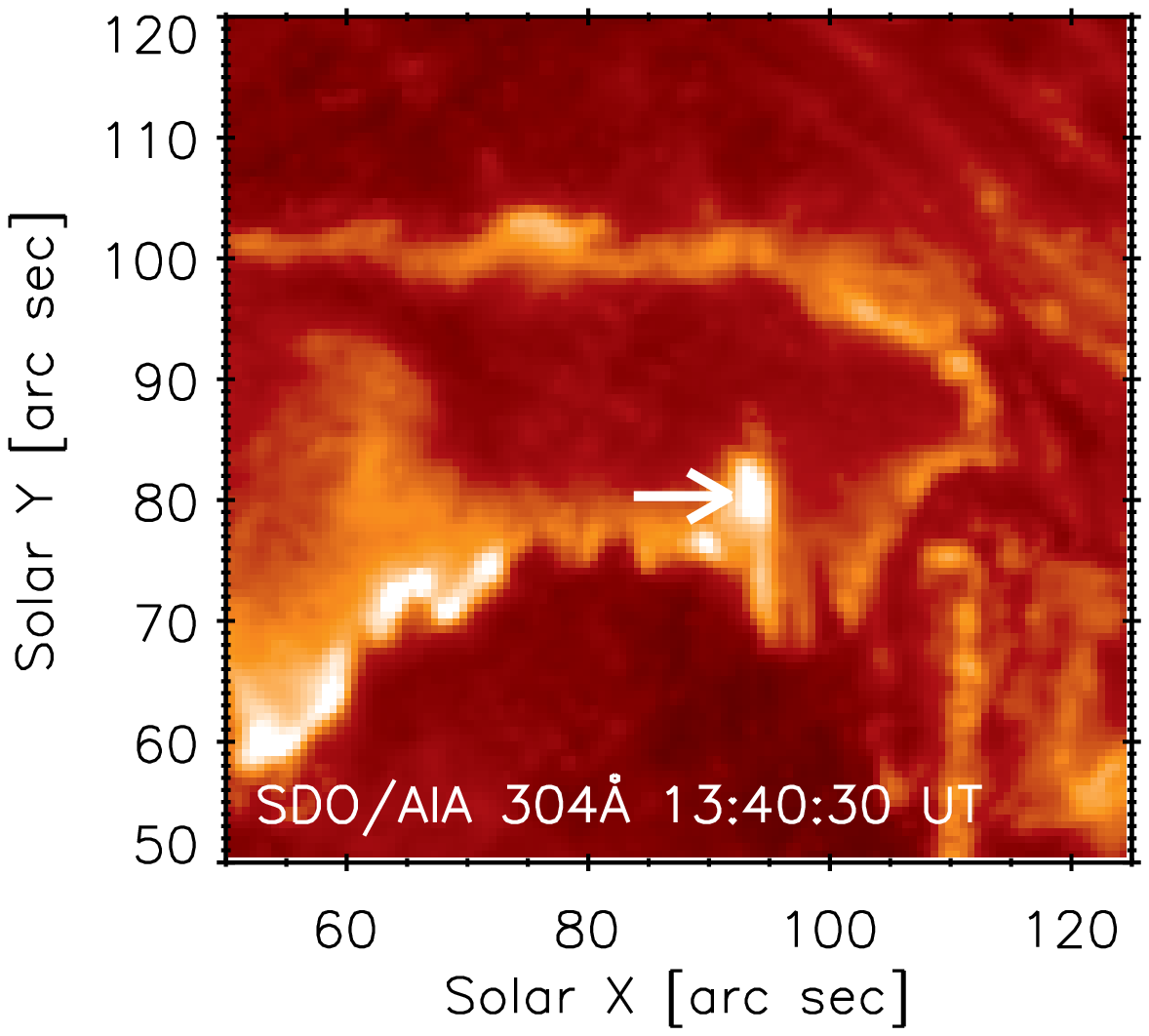}
        \includegraphics[width=4.11cm,clip,viewport=77  0 350 305]{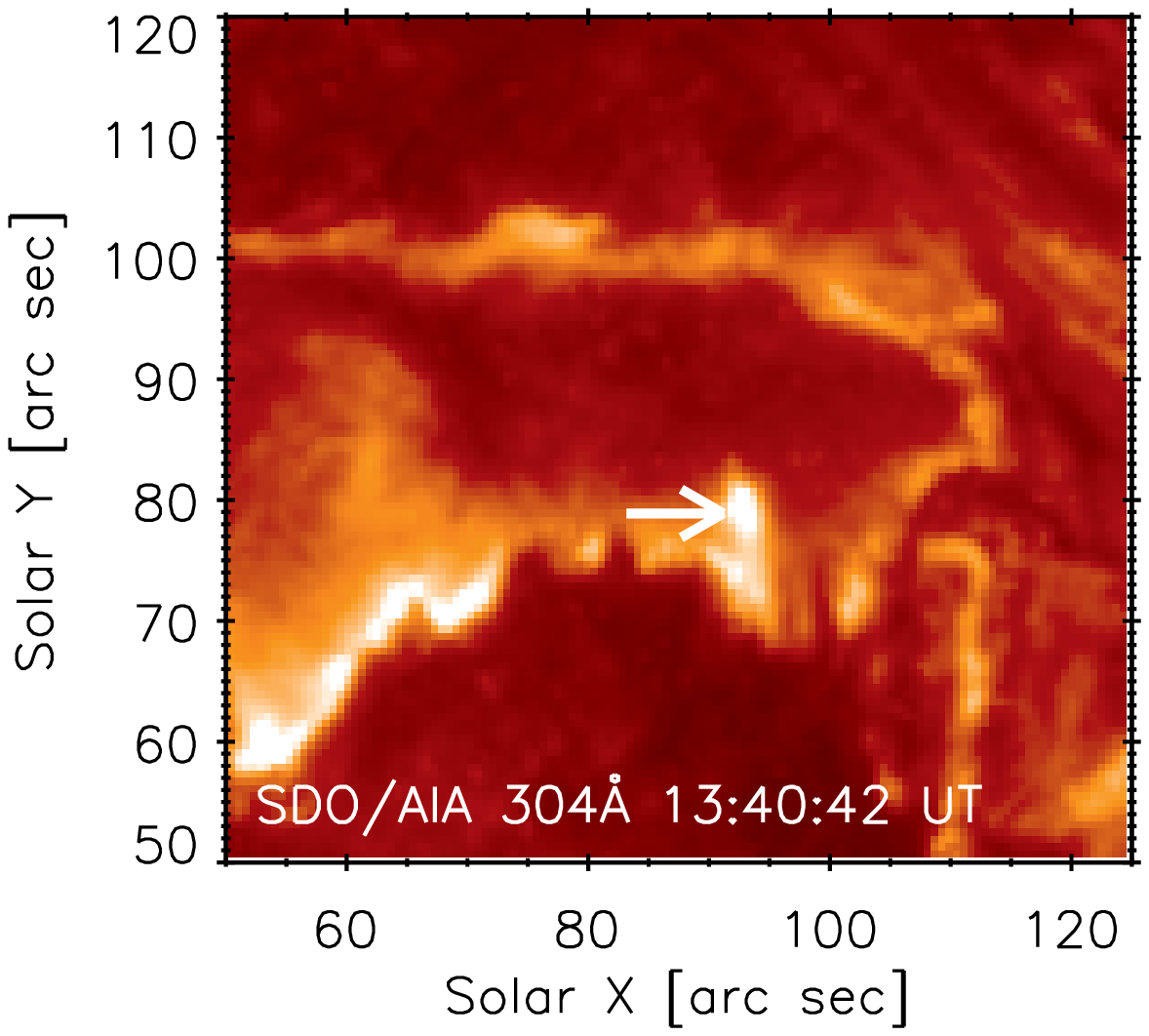}

\caption{Motion of plasma blob along the dark loop structure as seen in IRIS
1400\,\AA~slit jaw images and SDO/AIA 211\,\AA~filter (top row) and in SDO/AIA
304\,\AA~filter (bottom row). White arrows track the motion of the blob and
yellow ones point to the dark loop structure. The animation of AIA 304 filter
is available.} \label{fig4}
\end{figure*}

\begin{figure*}
        \centering
        \includegraphics[width=6.5cm,clip,viewport=0  0 160 130]{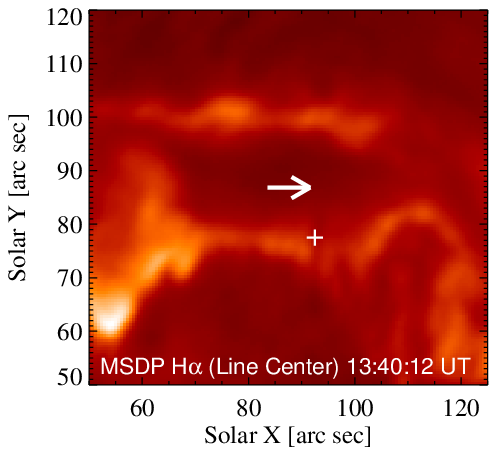}
        \includegraphics[width=5.5cm,clip,viewport=25  0 160 130]{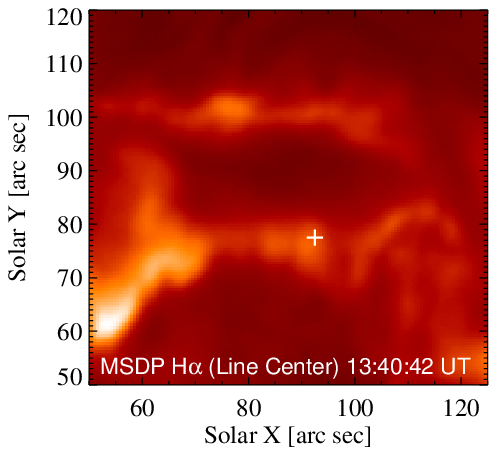}
        \includegraphics[width=5.6cm,clip,viewport=0  8 240 248]{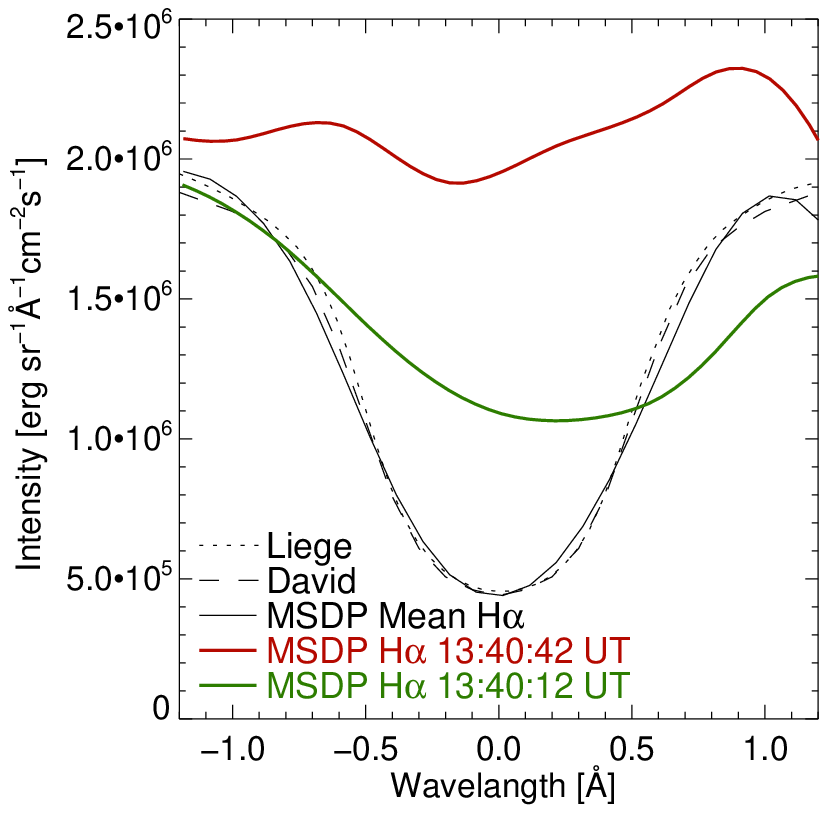}

\caption{MSDP H$_\alpha$ observations of M3.7 flare taken at 13:40:12~UT
(left), and 13:40:42~UT (middle). FOV is the same as in Figures 3 and 4. The
white arrow shows the position of the blob in SDO/AIA 94\AA~at 13:40:12 UT. The
last panel shows H$_\alpha$ line profiles taken at the same location marked by
white crosses in previous panels: green is the H$_\alpha$ line profile at
13:40:12~UT and red one at 13:40:42~UT. Solid black line presents H$_\alpha$
mean profile for quiet chromosphere from MSDP observations, dotted and dashed
lines are reference profiles from Liege and David catalogues, respectively. The
animation of H$_\alpha$ is available.} \label{fig5}
\end{figure*}

In such an interpretation, the velocity of the thermal front corresponds to the
the velocity of the falling blob and also to the sound speed of the hot plasma
at the bottom edge of the falling blob. If we take the velocity of the falling
blob (280 km s$^{-1}$) as the sound velocity, the corresponding plasma
temperature is about 3.5 MK (log($T$ [K])\,$\approx$\,6.54.) The bright blob
was observed in a broad range of temperatures from those in the transition
region through coronal ones up to flare temperatures. But this temperature is
lower than the temperature of the EUV lines dominating in 94\,\AA~and
131\,\AA~filters ($\approx$ 7 MK, i.e., log($T$ [K])\,$\approx$ 6.85).
According to the theory the velocity of the thermal front corresponds to the
sound velocity of the hot plasma close to the thermal front. Here the
temperature is lower than that in central part of the bright blob, because the
most energetic electrons from this region freely escape through the thermal
front to the region of cold plasma and propagate towards the chromosphere. Just
the flux of these energetic electrons can explain the maximum peak in the
derivatives of \textit{GOES} 1-8 and 0.5-4 \AA~soft X-rays (peak in hard X-ray
emission considering the Neupert effect) (Figure~\ref{fig1} b) at the time of
SPDB observation.

SPDB drifted from 1300 MHz to 2000 MHz during about 6 seconds. If in accordance
with the assumed plasma emission mechanism of SPDB we assume that the radio
frequencies correspond to the plasma frequencies, it means that the electron
plasma density in front of the thermal front changed from 2.08 $\times$
10$^{10}$ cm$^{-3}$ to 4.94 $\times$ 10$^{10}$ cm$^{-3}$ during 6 seconds.

Because there is no reliable density model for this case, in the following we
consider the density dependance in the gravitationally stratified solar
atmosphere: $n = n_0 \exp (-h/H)$, where $h$ is the height in the solar
atmosphere and $H$ is the scale height (H[m] = 50 T[K]). The hot blob with the
velocity 280 km s$^{-1}$ moves distance 1680 km during 6 seconds. If we
assume that this motion is in vertical direction, we can estimate
the scale height as $H$ = 1950 km, which gives the temperature in front of the
thermal conduction front (in front of the falling blob) 39000 K. However, the
blob falls along the cold loop structure, which is not vertical, therefore the
scale height is probably shorter and temperature is smaller and thus to be in
agreement with observations of the H$_\alpha$ cold loop (the temperature up to
about 20000 K). However, we cannot exclude that the density gradient in this
cold loop differs from that given by the gravitational equilibrium because
plasma processes at this stage are very dynamic.

In the region of the thermal front the plasma waves were generated and then
converted to the electromagnetic (radio) waves observed as SPDB. In this
complex process there are several possibilities how to explain the sudden start
and fading of SPDB. The most probable reason for this sudden start is sudden
formation of the thermal front. Because the collisional optical depth for the
plasma emission process increases as the second power of the radio frequency
\citep{Benz93}, the fading of SPDB towards higher frequencies can be explained
by this optical depth increase. Nevertheless, there are other possibilities
like a change of the level of the plasma waves and so on. In this emission
process the so called ducting \citep{Benz93} probably plays a role, because the
blob has a multithermal structure.

Just before the analyzed falling blob (which was associated with SPDB) we
observed other falling blobs with no radio counterparts. But they were less
bright and not visible in XRT Be\_med and Al\_thick filters. Note that for more
bright and hotter blob the association with the radio burst and also formation
of the thermal front is more probable. We think that in other cases maybe no
thermal fronts were formed or the radio emission was very weak, because this
emission is very sensitive to plasma conditions in the thermal front.

\acknowledgements We acknowledge support from Grants 16-13277S and 17-16447S of
the Grant Agency of the Czech Republic. K.R. was supported by the NCN grant no.
UMO-2015/17/B/ST9/02073.

\newpage

\end{document}